\documentclass[11pt,english]{article}
\usepackage[latin9]{inputenc}
\usepackage{amsmath}
\usepackage{amssymb}
\usepackage{graphicx}
\usepackage{setspace}
\makeatletter

\providecommand{\tabularnewline}{\\}

\numberwithin{equation}{section}

\@ifundefined{date}{}{\date{}}
\usepackage{esint}
\usepackage{hyperref}
\usepackage{latexsym}
\usepackage{graphicx}\usepackage{bm}\usepackage{longtable}

\usepackage{xcolor}

\@addtoreset{equation}{section}

\makeatother

\usepackage{babel}
\begin{document}
\title{Electromagnetic instability of vacuum with instantons in the holographic
plasma }
\maketitle
\begin{center}
Shao-cheng Hou\footnote{Email: housc@dlmu.edu.cn}, Si-wen Li\footnote{Email: siwenli@dlmu.edu.cn}, 
\par\end{center}

\vspace{4mm}

\begin{center}
\emph{School of Science, Dalian Maritime University, Dalian 116026,
China}\\
\par\end{center}

\vspace{12mm}

\begin{abstract}
Using the gauge-gravity duality, we study the electromagnetic instability
of vacuum with instantons in holographic plasma. The model we employ
is the D(-1)-D3 brane system in which the D(-1)-branes correspond
to the instantons in holography. To take into account the flavored
quarks, the coincident probe D7-branes as flavors are embedded into
the bulk geometry so that the effective electromagnetic Lagrangian
with flavors corresponds to the action of the D7-branes according
to gauge-gravity duality. We numerically evaluate the vacuum decay
rate, the critical electric field and the V-A curve of the vacuum
by using the D7-brane action with various values of the electromagnetic
field. It implies the particles in the plasma acquire an effective
mass in the presence of instantons as it is expected in the quantum
field theory, and the plasma trends to become insulating when the
electric field is small. This work reveals the relation between electromagnetic
and instantonic properties of the vacuum in the plasma.

\newpage{}
\end{abstract}
\tableofcontents{}

\section{Introduction}

Instanton in quantum chromodynamics (QCD) is known as the non-trivially
topological excitation of the vacuum. It relates to the breaking of
chiral symmetry contributing to the thermodynamics of QCD \cite{Gross:1980br,Schafer:1996wv,Vicari:2008jw}.
There have been a lot of works to connect the breaking of chiral symmetry
and confinement to the instanton constituents \cite{Diakonov:2004jn,Diakonov:2007nv,Liu:2015ufa,Liu:2015jsa,Liu:2016thw,Liu:2016mrk,Liu:2016yij,Poppitz:2012nz,Poppitz:2012sw}
since instanton is known to be consisted of BPS (Bogomol'nyi-Prasad-Sommerfield)
monopoles or dyons \cite{Tong:2005un}. And recently since the CP
violation in the decays of baryon is observed \cite{HFLAV:2022esi,LHCb:2025ray},
the breaking of chiral symmetry caused by QCD instanton again attracts
many interests in theory. 

However, using the perturbative method in quantum field theory (QFT)
to investigate QCD with instanton in the low-energy region is very
challenging since QCD is strongly coupled in this region due to its
property of asymptotic freedom. Fortunately, the gauge-gravity duality,
as an alternative method, provides us a holographic way to study the
strongly coupled QFT analytically \cite{Maldacena:1997re,Aharony:1999ti,Witten:1998qj}.
While the original version of the gauge-gravity duality illustrates
the equivalence between the IIB supergravity and $\mathcal{N}=4$
super Yang-Mills theory on the D3-branes, it is possible to include
the Yang-Mills instanton in this framework which corresponds to the
D(-1)-brane in the IIB string theory \cite{Liu:1999fc,Casalderrey-Solana:2011dxg}.
Accordingly, various properties of QCD with instantons are explored
by using the D(-1)-D3 brane system which is regarded as the holographic
researches of the literatures \cite{Gross:1980br,Schafer:1996wv,Vicari:2008jw},
e.g. chiral transition \cite{Gwak:2012ht}, heavy quark potential\cite{Chen:2022obe},
real time dynamics \cite{Li:2017ywp}, baryon spectrum and baryon
decay \cite{Li:2025ahp,Li:2015kma,Li:2015oza,Li:2016kfo,Li:2024apc},
thermodynamics and the topological properties of instantons \cite{Li:2021vve,Li:2022wwv}. 

Keeping the above in hand, in this work we would like to focus on
the electromagnetic instability in the holographic plasma with instantons.
The motivation is as follows. First, in the heavy-ion collision experiments,
since the charged particles move at very high speed, extremely strong
electromagnetic field would be generated at the collision. At this
moment, the virtual particles in the vacuum are possibly excited by
the electromagnetic field to be real particles. And it is known as
the Schwinger effect \cite{Schwinger:1951nm,Affleck:1981bma} which
is non-perturbative. Second, the QCD instanton is expected to affect
the process of particle creation through the Schwinger effect \cite{Buckley:1999mv}
which leads to observable results. In this sense, the concerned D(-1)-D3
brane system serves as the exact model to study the electromagnetic
instability in holography. To take into account the flavored quarks,
we can further introduce the coincident probe D7-branes as flavors
embedded into the bulk geometry produced by D(-1)- and D3-branes.
According to the dictionary of the gauge-gravity duality, the effective
flavored Lagrangian in the dual theory corresponds to the action of
the flavor brane, thus it is possible to evaluate the amplitude of
the vacuum decay by using $\left\langle 0\left|U\left(t\right)\right|0\right\rangle \sim e^{i\int\mathcal{L}d^{4}x}$
where $U\left(t\right)$ is the time-evolution operator. So the imaginary
part of the Lagrangian $\mathcal{L}$ is related to the vacuum decay
rate \cite{Rotter:2009zhs,Ge:2019crj,Shen:2018cjc}\footnote{The readers may review this holographic approaches without instantons
in \cite{Hashimoto:2013mua,Hashimoto:2014dza}.}. Our numerical calculation displays vacuum decay rate has a maximum
value for the given external electromagnetic field and the critical
electric field increases rapidly as the instanton density grows. The
V-A curve also confirm this feature which in addition implies the
vacuum has an insulating/conductive phase transition with respect
to the instanton density. The reason is that, in the presence of the
instantons, particles, e.g. quarks or mesons, in the plasma acquire
the effective mass from the chiral condensate, so the vacuum decay
can occur through a tunneling process when the electromagnetic field
strength is small. This implies the vacuum trends to be insulating
with instantons while it seems conductive without instantons. And
it agrees with the analysis of the electromagnetic features by using
a fundamental string in the D(-1)-D3 brane system \cite{Li:2020azb,Shahkarami:2015qff}.
Therefore our results reveal the relation between the instantonic
and conductive properties of the plasma.

The outline of this work is as follows. In Section 2, we review the
supergravity solution of the D(-1)-D3 brane system and the embedding
of the D7-branes as flavors briefly. In Section 3, we derive the effective
Lagrangian to describe the electromagnetic properties of the vacuum
with instantons. In Section 3, the vacuum decay rate and the V-A curve
are numerically evaluated. The summary of this work is given in the
final section.

\section{The holographic setup}

In this section, we will briefly review the supergravity background
for holographic plasma with instanton which is constructed in the
framework of type IIB string theory by using $N_{c}$ D3-branes with
$N_{\mathrm{D}}$ D-instantons i.e. the D(-1)-branes in the large-$N_{c}$
limit \cite{Liu:1999fc}. And this system describes instantonic plasma
at finite temperature. In addition, we will introduce $N_{f}$ D7-branes
into the D(-1)-D3-brane background as flavors \cite{Gwak:2012ht},
hence the energy of flavor vacuum can be obtained by evaluating the
classical action of the D7-branes, which is useful to study the vacuum
instability with electromagnetic field.

\subsection{The D3-brane background with D-instanton}

We start with the IIB supergravity action which describes the dynamics
of $N_{c}$ black D3-branes with $N_{\mathrm{D}}$ D-instantons in
the large $N_{c}$ limit. This system is recognized as a marginal
\textquotedblleft bound state\textquotedblright{} of D3-branes with
smeared $N_{\mathrm{D}}$ D(-1)-branes. In string frame, the supergravity
action is given as, 

\begin{equation}
S_{\mathrm{IIB}}=\frac{1}{2\kappa_{10}^{2}}\int d^{10}x\sqrt{-g}\left[e^{-2\Phi}\left(\mathcal{R}+4\partial\Phi\cdot\partial\Phi\right)-\frac{1}{2}\left|F_{1}\right|^{2}-\frac{1}{2}\left|F_{5}\right|^{2}\right],\label{eq:2.1}
\end{equation}
 where $l_{s}$ is the string length, $2\kappa_{10}^{2}=\left(2\pi\right)^{7}l_{s}^{8}$
refers to the 10d gravity coupling constant. We use $\Phi$ to refer
to the dilaton field and use $F_{1,5}$ to denote the field strength
of the Ramond-Ramond (R-R) zero and four form $C_{0,4}$ respectively.
Note that, $N_{c}$ D3-branes are identified as the color branes.
In the near-horizon region, the geometric background of $N_{c}$ black
D3-branes with $N_{\mathrm{D}}$ D-instantons is the solution of non-extremal
D3-branes with a non-trivial $C_{0}$, it reads \cite{Gwak:2012ht},

\begin{align}
ds^{2} & =e^{\frac{\phi}{2}}\left\{ \frac{R^{2}}{z^{2}}\left[-f\left(z\right)dt^{2}+d\mathbf{x}\cdot d\mathbf{x}+\frac{dz^{2}}{f\left(z\right)}\right]+R^{2}d\Omega_{5}^{2}\right\} ,\nonumber \\
e^{\phi} & =1-z_{H}^{4}q\ln f\left(z\right),\ f\left(r\right)=1-\frac{z^{4}}{z_{H}^{4}},\ F_{5}=dC_{4}=g_{s}^{-1}\mathcal{Q}_{3}\epsilon_{5},\nonumber \\
F_{1} & =dC_{0},\ C_{0}=-ie^{-\phi}+i\mathcal{C},\ \phi=\Phi-\Phi_{0},\ e^{\Phi_{0}}=g_{s},\label{eq:2.2}
\end{align}
where $\epsilon_{5}$ is the volume element of a unit $S^{5}$, $\mathcal{C}$
is a boundary constant and $g_{s}$ is the string coupling constant.
And the presented parameters are given as,

\begin{equation}
R^{4}=4\pi g_{s}N_{c}l_{s}^{4},\ \mathcal{Q}_{3}=4R^{4},\ Q=\frac{N_{\mathrm{D}}}{N_{c}}\frac{\left(2\pi\right)^{4}\alpha^{\prime2}}{V_{4}}\mathcal{Q}_{3},q=\frac{Q}{R^{8}}
\end{equation}
In our notation, coordinates of $x^{\mu}=\left\{ t,\mathbf{x}\right\} =\left\{ t,x^{i}\right\} ,\ i=1,2,3$
denote the 4d spacetime $\mathbb{R}^{4}$ where the D3-branes are
extended along. The holographic direction perpendicular to the D3-branes
is denoted as $z$ and the holographic boundary is located at $z=0$.
The solution (\ref{eq:2.2}) is asymptotic $\mathrm{AdS}_{5}\times S^{5}$
at $z\rightarrow0$ which illustrates that D-instanton charge $N_{\mathrm{D}}$
is smeared homogeneously over the worldvolume $V_{4}$ of the $N_{c}$
black D3-branes with a horizon at $z=z_{H}$. Note that the ratio
$N_{\mathrm{D}}/N_{c}$ must be fixed in the large-$N_{c}$ limit,
since the backreaction of the D-instantons has been taken into account
in the background. So the dual theory of this background is conjectured
as the 4d $\mathcal{N}=4$ super Yang-Mills theory in a self-dual
gauge field (instantonic) background at finite temperature \cite{Liu:1999fc,Casalderrey-Solana:2011dxg}.
Besides, the R-R field $C_{0}$ is recognized as axion field in terms
of hadron physics and the gluon condensate (or chiral condensate)
in this system is evaluated as,

\begin{equation}
\left\langle \mathrm{Tr}F_{\mu\nu}F^{\mu\nu}\right\rangle \simeq\frac{N_{\mathrm{D}}}{16\pi^{2}V_{4}}=\frac{Q}{\left(2\pi\alpha^{\prime}\right)^{2}R^{4}}\frac{N_{c}}{\left(2\pi\right)^{4}}\propto\left\langle \bar{q}q\right\rangle ,\ \mu,\nu=0,1...3.\label{eq:2.4}
\end{equation}
And the dual theory can be examined by introducing a probe D3-brane
located at the holographic boundary $z=0$ whose bosonic action is,

\begin{align}
S_{\mathrm{D}3} & =\left[-T_{\mathrm{D}3}\int d^{4}xe^{-\frac{\phi}{2}}\mathrm{Str}\sqrt{-\det\left(g+\mathcal{F}\right)}+T_{\mathrm{D}3}\int C_{4}+\frac{1}{2}T_{\mathrm{D}3}\mathrm{Tr}\int C_{0}\mathcal{F}\land\mathcal{F}\right]\bigg|_{r\rightarrow\infty}\nonumber \\
 & \simeq-\frac{1}{4g_{\mathrm{YM}}^{2}}\int d^{4}xF_{\mu\nu}F^{\mu\nu}+\frac{\kappa}{2}\mathrm{Tr}\int F\wedge F+\mathcal{O}\left(F^{3}\right).
\end{align}
Here $\mathcal{F}=2\pi\alpha^{\prime}F$ refers to the gauge field
strength on the D3-brane and $T_{\mathrm{D3}}=g_{s}^{-1}\left(2\pi\right)^{-3}l_{s}^{-4}$
denotes the tension of D3-brane. $\kappa$ is a constant given by
the integral of $C_{0}$ at boundary and $g_{\mathrm{YM}}$ refers
to the Yang-Mills coupling constant in the dual theory. Hence we can
see the dual theory to the D(-1)-D3-brane system is $\mathcal{N}=4$
super Yang-Mills theory with a self-dual gauge field, or equivalently
with axion or theta term.

\subsection{The D7-branes as flavors and the electromagnetic instability}

In order to introduce the flavored fermions as hypermultiplet, the
D7-brane as flavors is necessary as it is discussed in the D3/D7 approach
\cite{Karch:2002sh}. In this work, as the concern is the instability
induced by electromagnetic field, we consider the massless hypermultiplet
since, in the side of QFT, instanton does not lead to additional vacuum
instability for massless fermion \cite{Gross:1980br,Schafer:1996wv,Vicari:2008jw}.
The configuration of the D-branes in this system is given in Table
\ref{tab:1}. 
\begin{table}
\begin{centering}
\begin{tabular}{|c|c|c|c|c|c|c|c|c|c|c|c|}
\hline 
 & (-1) & 0 & 1 & 2 & 3 & 4 & 5 & 6 & 7 & 8 & 9\tabularnewline
\hline 
\hline 
D(-1) & - &  &  &  &  &  &  &  &  &  & \tabularnewline
\hline 
D3-brane &  & - & - & - & - &  &  &  &  &  & \tabularnewline
\hline 
D7-brane &  & - & - & - & - & - & - & - & - &  & \tabularnewline
\hline 
\end{tabular}
\par\end{centering}
\caption{\label{tab:1} The configuration of the D-branes in the D(-1)-D3-brane
system.}
\end{table}
 The flavored hypermultiplet comes from oscillations of the 3-7 string
which refers to a string connecting the D3- and the D7-branes, therefore
the configuration of D7-brane touching the stack of D3-branes illustrates
the massless-ness of the hypermultiplet. Since we will focus on the
electromagnetic instability of the flavored vacuum, the bosonic part
of a single D7-brane action is needed as,

\begin{equation}
S_{\mathrm{D}7}=-T_{\mathrm{D7}}\int dtd^{3}xdzd\Omega_{3}e^{-\phi}\sqrt{-\det\left(g_{MN}+2\pi\alpha^{\prime}F_{MN}\right)},\label{eq:2.6}
\end{equation}
where the indices $M,N$ run over the D7-brane, $T_{\mathrm{D7}}=g_{s}^{-1}\left(2\pi\right)^{-7}l_{s}^{-8}$
refers to the tension of D7-brane, $g_{MN}$ is the induced metric
on the D7-brane and $F_{MN}$ refers to the gauge field strength on
the D7-brane worldvolume. The integration measure $dz,d\Omega_{3}$
refers to the extra four dimensions of the D7-brane which are vertical
to the worldvolume of D3-brane as 1+3 spacetime of $\mathbb{R}^{4}$.

In QFT, the electromagnetic instability and vacuum decay rate can
be evaluated by analyzing the effective Lagrangian $\mathcal{L}$
since it relates to the vacuum-to-vacuum amplitude \cite{Hashimoto:2013mua,Hashimoto:2014dza}
as,

\begin{equation}
\left\langle 0\left|U\left(t\right)\right|0\right\rangle =e^{i\mathcal{L}vt},
\end{equation}
where $U\left(t\right)$ is the time-evolution operator with external
electromagnetic fields, $v$ denotes the spatial volume and $\left|0\right\rangle $
represents the vacuum state without any external fields. In particular,
considering the AdS/CFT dictionary in our setup, the effective Lagrangian
$\mathcal{L}$ must be able to describe the vacuum dynamics of flavors
which is expected to be (\ref{eq:2.6}) in holography. In this sense,
if the effective Lagrangian (\ref{eq:2.6}) has an imaginary part
$\Gamma$ as \footnote{For example, the Lagrangian of QED has the imaginary part up to 1-loop
order as, 
\begin{align*}
\mathrm{Im}\mathcal{L}_{\mathrm{spinor}}^{\mathrm{1-loop}} & =\frac{e^{2}E^{2}}{8\pi^{3}}\sum_{n=1}^{\infty}\frac{1}{n^{2}}\mathrm{exp}\left(-\frac{\pi m^{2}}{eE}n\right),\\
\mathrm{Im}\mathcal{L}_{\mathrm{scalar}}^{\mathrm{1-loop}} & =\frac{e^{2}E^{2}}{16\pi^{3}}\sum_{n=1}^{\infty}\frac{1}{n^{2}}\mathrm{exp}\left(-\frac{\pi m^{2}}{eE}n\right),
\end{align*}
which corresponds to a single quantum tunneling process in Schwinger
effect. It illustrates a pair of an electron and a positron is created
from the vacuum.},

\begin{equation}
\mathcal{L}=\mathrm{Re}\mathcal{L}+i\frac{\Gamma}{2},
\end{equation}
it could be interpreted as the the vacuum decay rate in holography. 

To investigate the vacuum decay rate $\Gamma$ with respect to the
electromagnetic instability, we can turn on the static gauge field
potential as $A_{\mu}=\left(A_{0},A_{1},0,0\right)$ with the gauge
condition $A_{z}=0$ for simplicity and it implies the electric field
can be fixed along $x^{1}$ due to the rotation symmetry. Besides,
the gauge field potentials are functions as $A_{0}\left(\mathbf{x},z\right),A_{1}\left(\mathbf{x},z\right)$
to give the electromagnetic fields $E_{i},B_{i}$ and include the
holographic information, and we further assume that the electromagnetic
fields $E_{i},B_{i}$ are constants as the external fields. Keep all
the above in hand, we can derive the effective Lagrangian $\mathcal{L}$
to evaluate the vacuum instability from (\ref{eq:2.6}) as\footnote{There should be a Wess-Zumino term in the total action for the D7-brane,
however it vanishes in our current setup.},

\begin{align}
S_{\mathrm{D}7} & =-\mu_{7}\int dtd^{3}xdzd\Omega_{3}e^{-\phi}\sqrt{-\det\left(g_{MN}+2\pi\alpha^{\prime}F_{MN}\right)}\nonumber \\
 & =-2\pi^{2}\mu_{7}V_{4}R^{8}\mathcal{L},\label{eq:2.9}
\end{align}
where

\begin{align}
\mathcal{L} & =\int_{z_{H}}^{0}\frac{e^{\phi}}{z^{5}}\sqrt{\xi},\nonumber \\
\xi & =1-\frac{\left(2\pi\alpha^{\prime}\right)^{2}z^{4}}{R^{4}}e^{-\phi}\left(F_{0z}^{2}+F_{01}^{2}f^{-1}-F_{1z}^{2}f-F_{12}^{2}-F_{23}^{2}-F_{13}^{2}\right)\nonumber \\
 & -\frac{\left(2\pi\alpha^{\prime}\right)^{4}z^{8}}{R^{8}}e^{-2\phi}\left[F_{23}^{2}\left(F_{01}^{2}f^{-1}-F_{1z}^{2}f\right)+F_{0z}^{2}\left(F_{12}^{2}+F_{23}^{2}+F_{13}^{2}\right)\right],\label{eq:2.10}
\end{align}
and $F_{01}^{2}=E_{1}^{2},F_{23}^{2}=B_{1}^{2},F_{12}^{2}+F_{23}^{2}+F_{13}^{2}=B_{3}^{2}+B_{1}^{2}+B_{2}^{2}$.
Therefore we can see $\xi$ could be negative in order to lead to
an imaginary part of $\mathcal{L}$, if the electromagnetic becomes
sufficiently large.

Varying the effective Lagrangian (\ref{eq:2.10}) with respect to
$A_{0}$ and $A_{1}$, the associated equations of motion for the
gauge field potential can be obtained as,

\begin{align}
\partial_{1}\frac{\partial\mathcal{L}}{\partial\left(\partial_{1}A_{0}\right)}+\partial_{z}\frac{\partial\mathcal{L}}{\partial\left(\partial_{z}A_{0}\right)} & =-\partial_{z}\left(\frac{R^{8}}{z^{5}}e^{\phi}\frac{1}{2\sqrt{\xi}}\frac{\partial\xi}{\partial F_{0z}}\right)=0,\nonumber \\
\partial_{i}\frac{\partial\mathcal{L}}{\partial\left(\partial_{i}A_{1}\right)}+\partial_{z}\frac{\partial\mathcal{L}}{\partial\left(\partial_{z}A_{1}\right)} & =-\partial_{z}\left(\frac{R^{8}}{z^{5}}e^{\phi}\frac{1}{2\sqrt{\xi}}\frac{\partial\xi}{\partial F_{1z}}\right)=0,
\end{align}
which leads to two constants as the electric charge $d$ and current
$j$ given as,

\begin{align}
d & =-\frac{R^{8}}{z^{5}}e^{\phi}\frac{1}{2\sqrt{\xi}}\frac{\partial\xi}{\partial F_{0z}}=\frac{2\pi\alpha^{\prime}}{z\sqrt{\xi}}\left[1+\frac{\left(2\pi\alpha^{\prime}\right)^{2}z^{4}}{R^{4}}e^{-\phi}\left(F_{12}^{2}+F_{23}^{2}+F_{13}^{2}\right)\right]F_{0z},\nonumber \\
j & =-\frac{R^{8}}{z^{5}}e^{\phi}\frac{1}{2\sqrt{\xi}}\frac{\partial\xi}{\partial F_{1z}}=\frac{2\pi\alpha^{\prime}}{z\sqrt{\xi}}\left[1+\frac{\left(2\pi\alpha^{\prime}\right)^{2}z^{4}}{R^{4}}e^{-\phi}F_{23}^{2}\right]F_{1z}f.\label{eq:2.12}
\end{align}
Plugging (\ref{eq:2.12}) back into (\ref{eq:2.10}), $\xi$ can be
rewritten in term as,

\begin{equation}
\xi=\frac{1-\frac{\left(2\pi\alpha^{\prime}\right)^{2}z^{4}}{R^{4}}e^{-\phi}\left(E_{1}f^{-1}-B^{2}\right)-\frac{\left(2\pi\alpha^{\prime}\right)^{4}z^{8}}{R^{8}}e^{-2\phi}E_{1}^{2}B_{1}^{2}f^{-1}}{1-\frac{z^{6}j^{2}f^{-1}}{e^{\phi}R^{4}+\left(2\pi\alpha^{\prime}\right)^{2}B_{1}^{2}z^{4}}+\frac{z^{6}d^{2}}{e^{\phi}R^{4}+\left(2\pi\alpha^{\prime}\right)^{2}B^{2}z^{4}}}.\label{eq:2.13}
\end{equation}

\section{Vacuum properties from the holographic Lagrangian}

\subsection{Vacuum decay rate and critical electric field}

\begin{figure}
\begin{centering}
\includegraphics[scale=0.35]{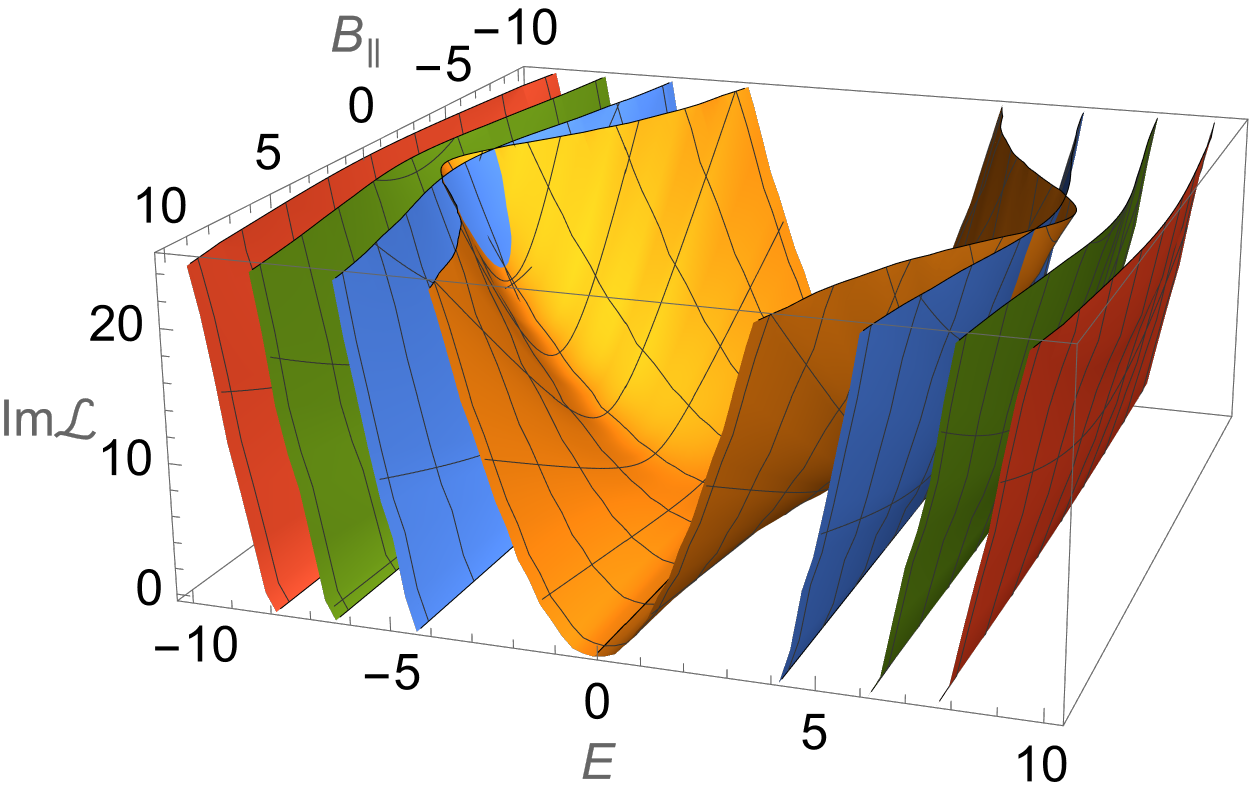}\includegraphics[scale=0.35]{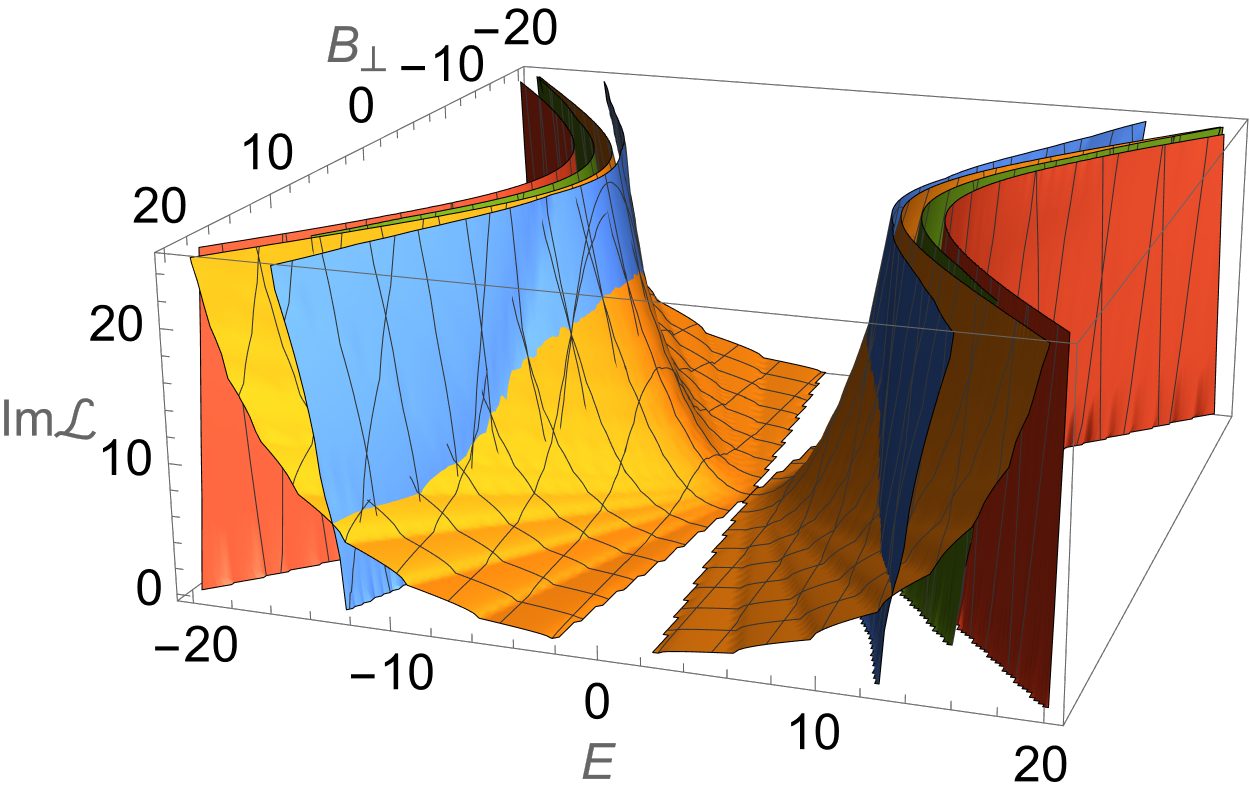}
\par\end{centering}
\caption{\label{fig:1}The imaginary part of effective Lagrangian as a function
of $E,B_{\parallel}$ and $E,B_{\perp}$ with various instanton charge
$q$. $B_{\parallel},B_{\perp}$ refers respectively to the cases
that the magnetic field is parallel and perpendicular to the electric
field. The parameters are chosen as $z_{H}=R=2\pi\alpha^{\prime}=1,d=0,j=0$.
The yellow, blue, green and red colors correspond respectively to
the case of $q=0,3,6,9$.}
\end{figure}
 
\begin{figure}
\begin{centering}
\includegraphics[scale=0.35]{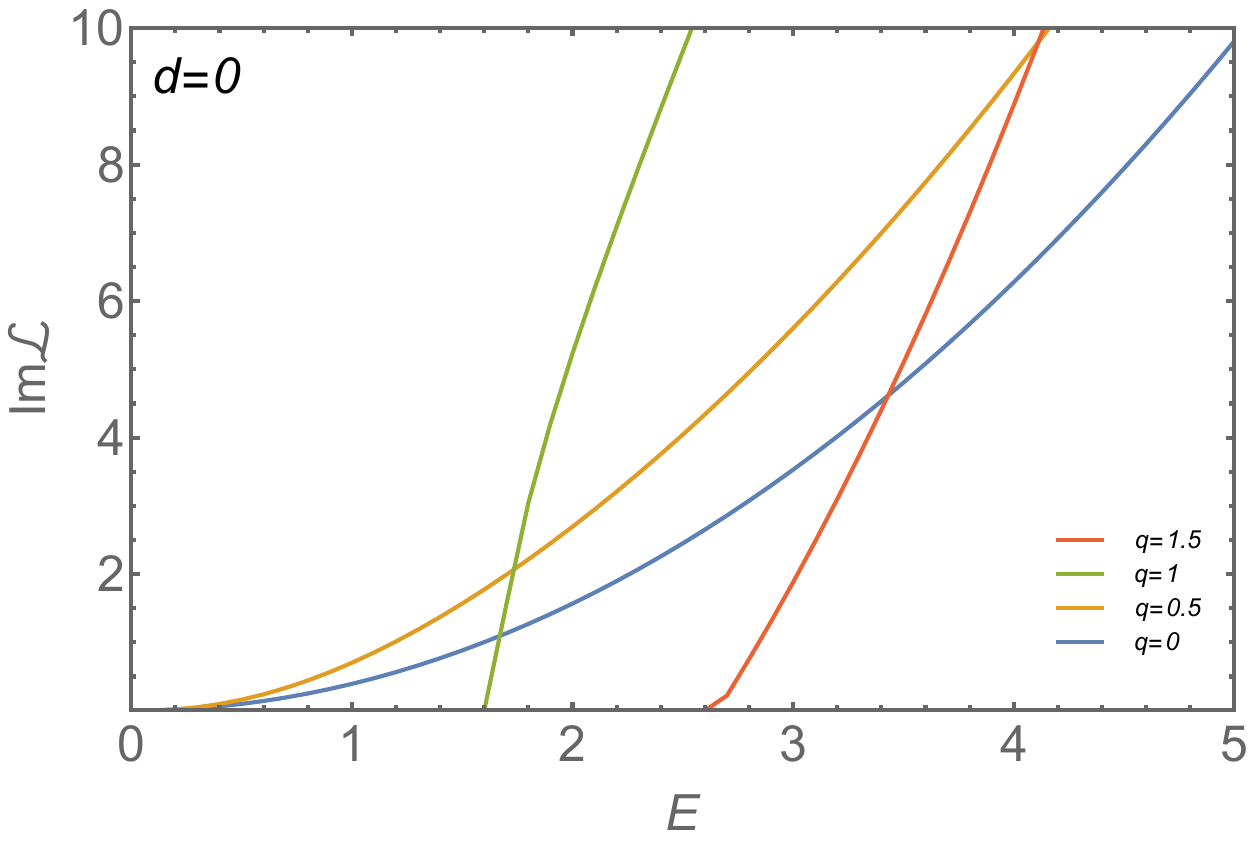}\includegraphics[scale=0.35]{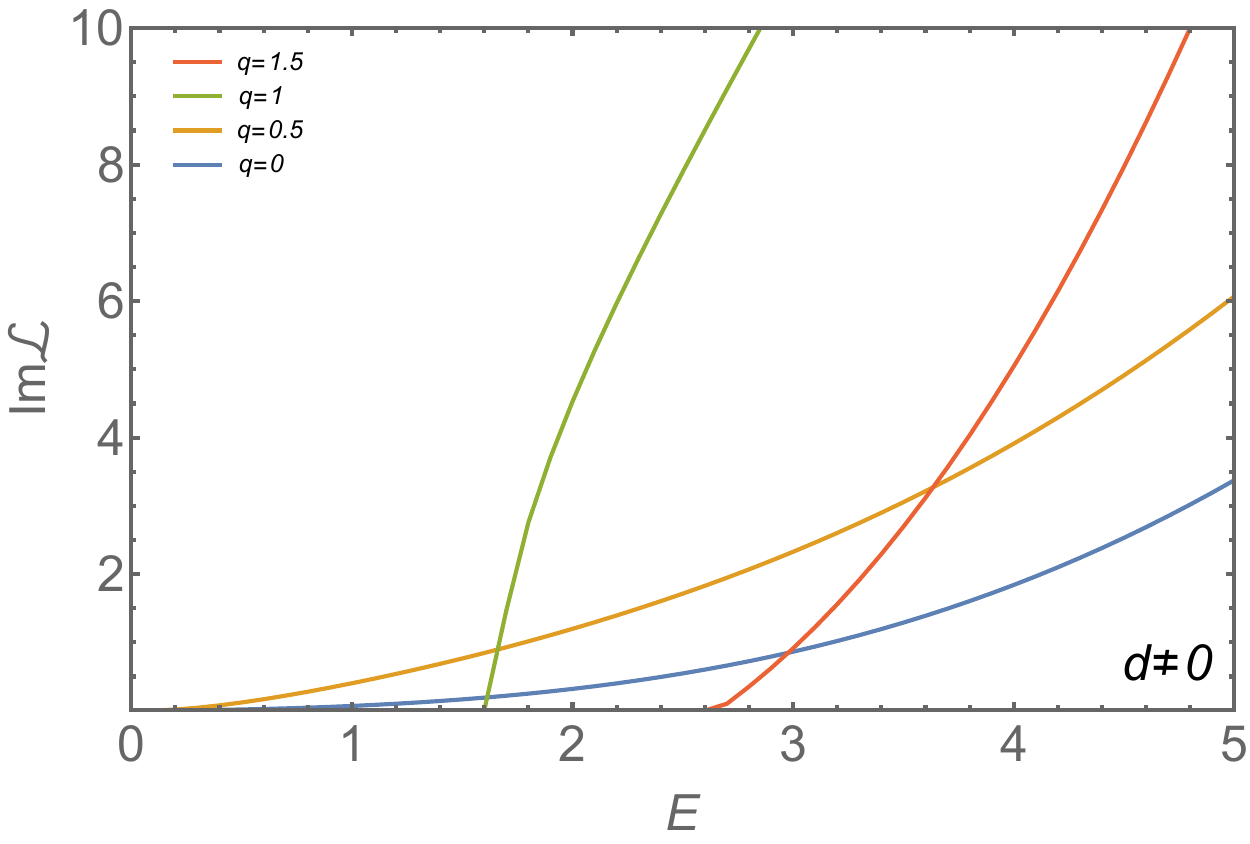}
\par\end{centering}
\caption{\label{fig:2}The imaginary part of effective Lagrangian as a function
of $E$ with various instanton charge $q$. The parameters are chosen
as $z_{H}=R=2\pi\alpha^{\prime}=1,j=0,B_{i}=0$.}
\end{figure}
\begin{figure}
\begin{centering}
\includegraphics[scale=0.35]{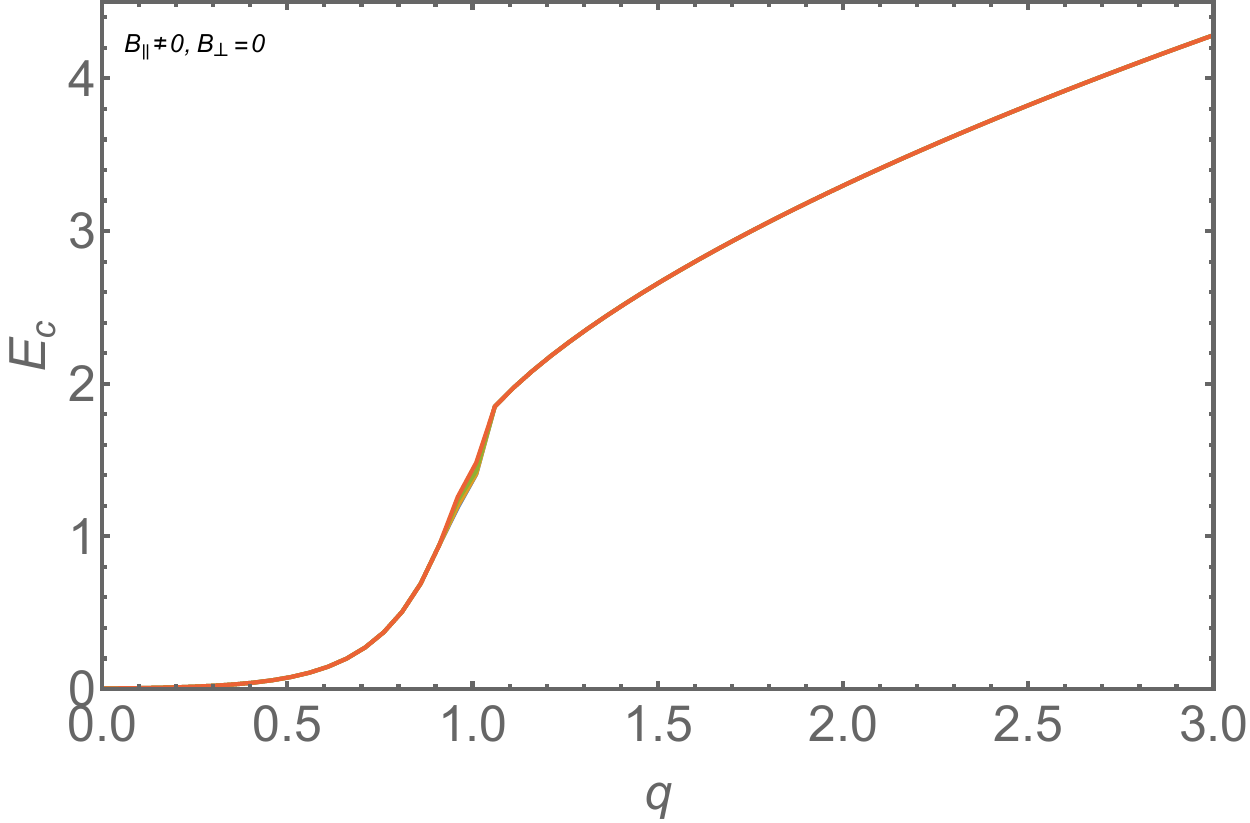}\includegraphics[scale=0.35]{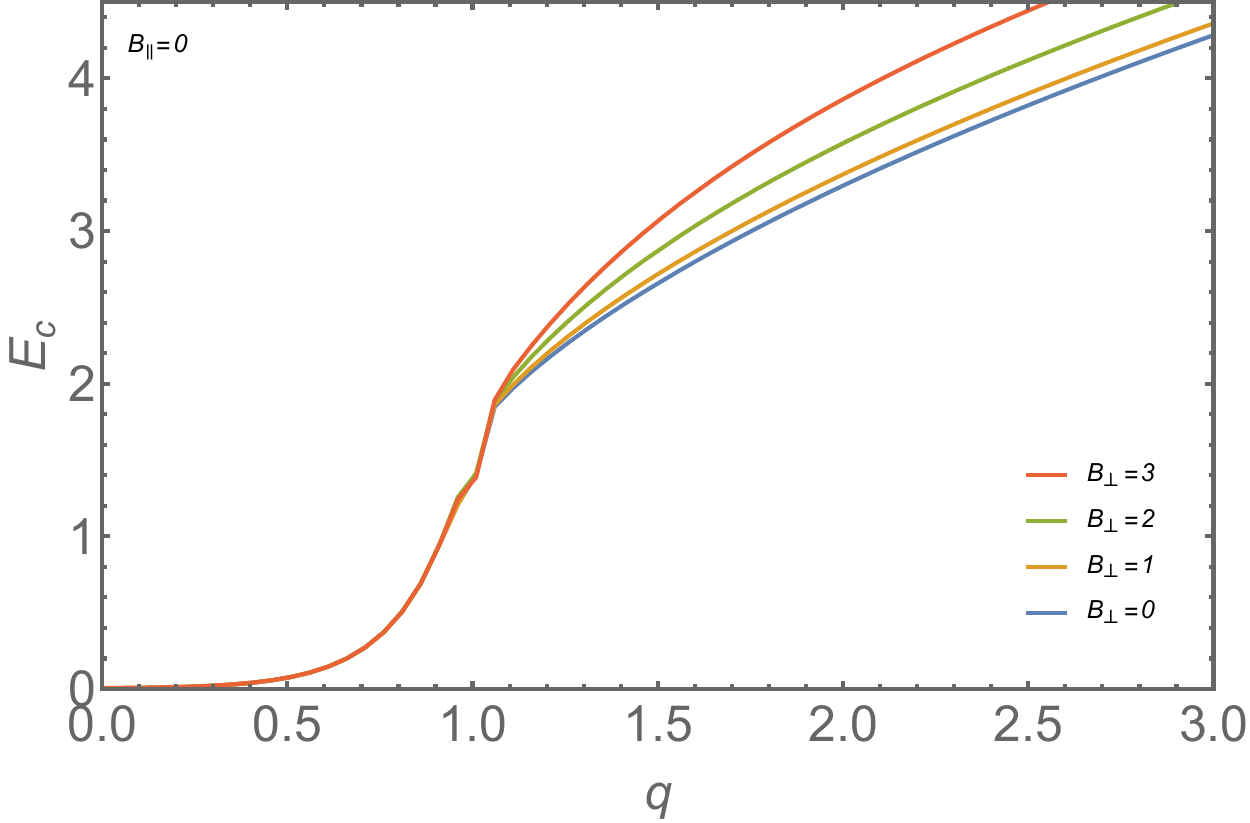}
\par\end{centering}
\caption{\label{fig:3}The critical electric field $E_{c}$ as a function of
the instanton charge $q$. The parameters are chosen as $z_{H}=R=2\pi\alpha^{\prime}=1,j=0$.}
\end{figure}

Since we focus on the vacuum decay here, the electric charge $d$
and current $j$ can be simply set to zero in the effective Lagrangian
(\ref{eq:2.9}) and (\ref{eq:2.13}) to evaluate the vacuum decay
rate. The associated numerical results are illustrated in Figure \ref{fig:1}
and Figure \ref{fig:2}. The numerical calculation reveals that the
electromagnetic instability of vacuum is very different with non-vanished
instanton charge denoted by $q$. Without the instanton charge, the
critical electric field trends to be zero which implies a very small
electric field can induce the vacuum decay. However, in the presence
of the instanton density $q>0$, it leads to a non-zero critical value
for the electric field. To clarify this, we further plot out the relation
between the critical electric field $E_{c}$ and the instanton charge
as it is illustrated in Figure \ref{fig:3} 
which demonstrates the critical electric field depends on the instanton
density quadratically for small $q$ and nearly linearly for large
$q$.

\begin{figure}[t]
\begin{centering}
\includegraphics[scale=0.37]{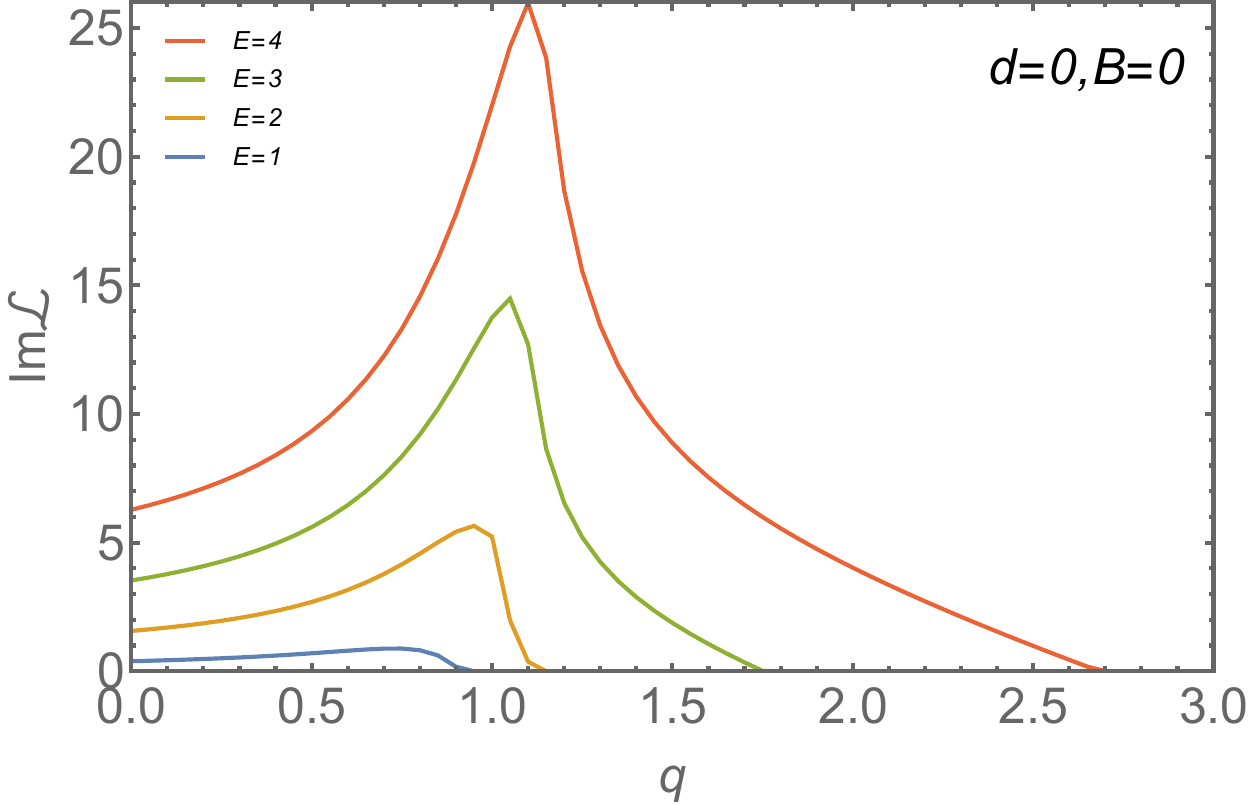}\includegraphics[scale=0.37]{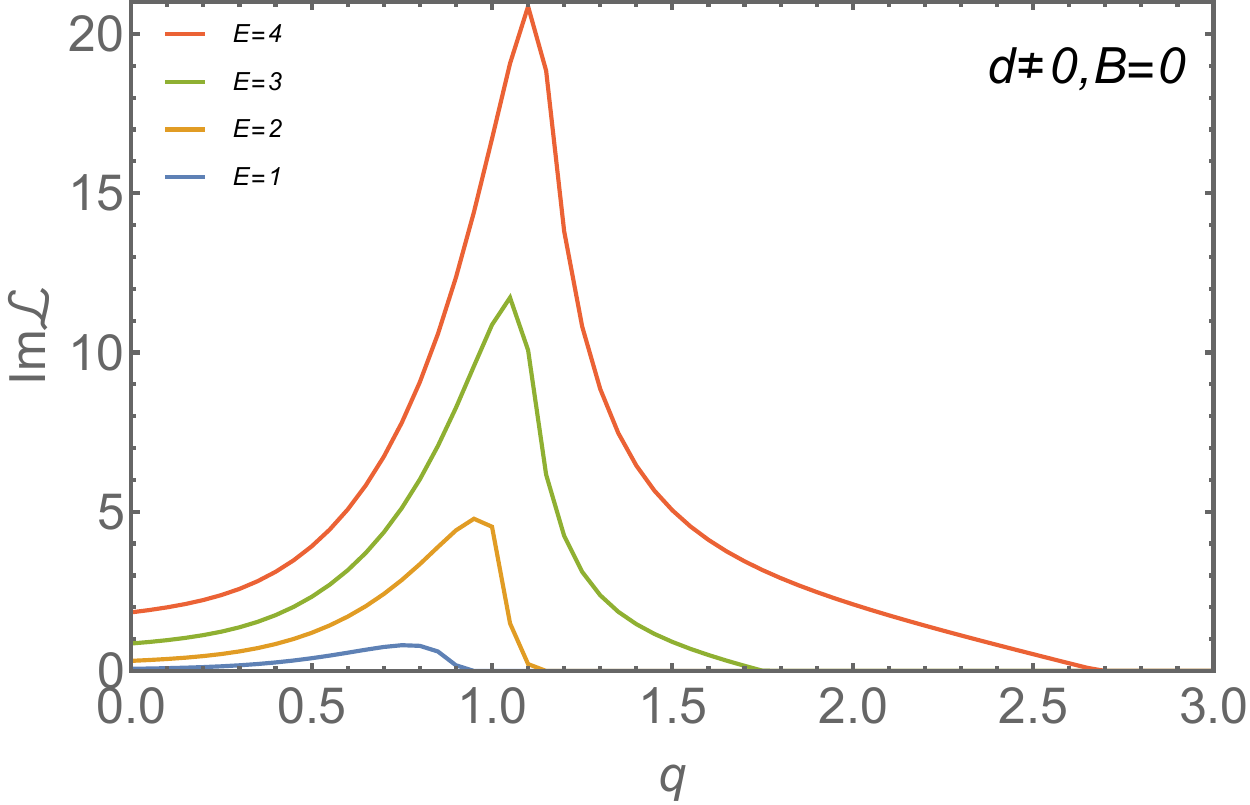}
\par\end{centering}
\begin{centering}
\includegraphics[scale=0.37]{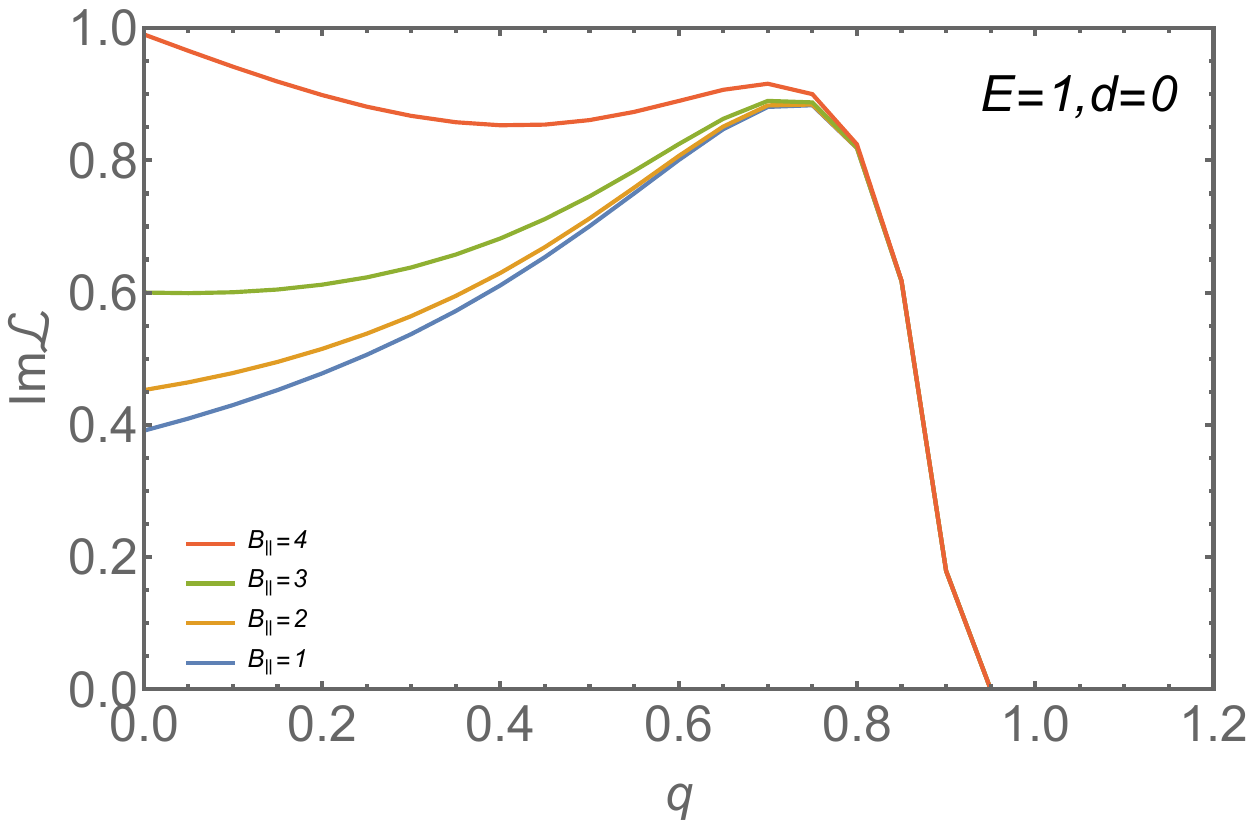}\includegraphics[scale=0.37]{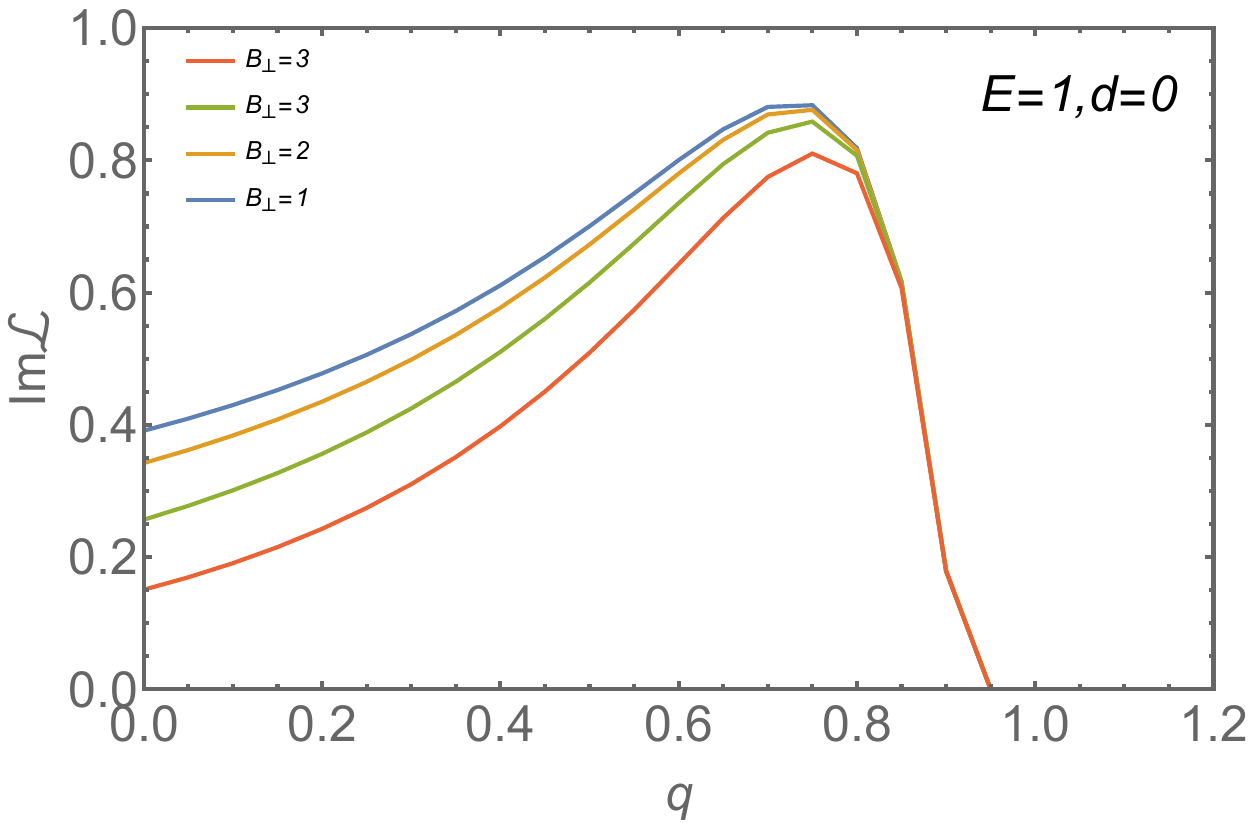}
\par\end{centering}
\caption{\label{fig:4}The imaginary part of effective Lagrangian as a function
of $q$ with fixed external fields. The parameters are chosen as $z_{H}=R=2\pi\alpha^{\prime}=1,j=0$.}
\end{figure}

To understand this behavior, let us introduce a probe D3-brane near
the holographic boundary at $z=z_{0}$ with an electric field $E_{1}$
as the most discussion about the holographic Schwinger effect \cite{Semenoff:2011ng}.
The action for a D3-brane is given as,

\begin{align}
S_{\mathrm{D3}} & =\int d^{4}xe^{-\frac{\phi}{2}}\sqrt{-\det\left(g+2\pi\alpha^{\prime}F\right)}\big|_{z=z_{0}}\nonumber \\
 & =\int d^{4}x\frac{R^{2}}{z}\sqrt{\frac{e^{\phi}fR^{4}}{z^{2}}-\left(2\pi\alpha^{\prime}\right)^{2}E_{1}^{2}}\big|_{z=z_{0}}.\label{eq:3.1}
\end{align}
Since the stable action requires that the square root presented in
(\ref{eq:3.1}) must be positive, it leads to a critical value $E_{c}$
of $E_{1}$ as,

\begin{equation}
E_{c}=\frac{e^{\frac{\phi\left(z_{0}\right)}{2}}\sqrt{f\left(z_{0}\right)}R^{2}}{2\pi\alpha^{\prime}z_{0}}.\label{eq:3.2}
\end{equation}
Recall the solution (\ref{eq:2.2}) for $\phi$ and expand (\ref{eq:3.2})
as series of $q$, we therefore find 

\begin{align}
E_{c} & \propto q^{2},\ q\ll1,\nonumber \\
E_{c} & \propto\sqrt{q},\ q\gg1,
\end{align}
which is nicely consistent with the numerical evaluation presented
in Figure \ref{fig:3}. This analysis also implies the vacuum decay
occurs only in a region with special values of $q$ when the electromagnetic
field is fixed as an external field. The numerical confirmation is
presented in Figure \ref{fig:4} 
in which the imaginary part of effective Lagrangian (\ref{eq:2.10})
is a function of $q$. Accordingly, we can see while electromagnetic
field increases the vacuum decay rate, the imaginary part of effective
Lagrangian is non-vanished only in a special region. 

In addition, the increase of the critical electric field in the presence
of instanton is due to the gluon condensate given (\ref{eq:2.4})
which is equivalently a potential in the vacuum Schwinger effect as
it is discussed in \cite{Li:2020azb,Shahkarami:2015qff,Semenoff:2011ng}.
Hence the Schwinger effect or vacuum decay will not occur if electric
field is less than this potential. That is why we find a non-zero
critical value of the electric field in the presence of the instanton.
Therefore, our holographic description agrees basically with the conclusion
that instanton affects the vacuum structure \cite{Li:2020azb,Shahkarami:2015qff}.
Notice that the magnetic field with vanished electric field does not
occur the vacuum decay according to (\ref{eq:2.13}).

\subsection{The V-A curve}

\begin{figure}
\begin{centering}
\includegraphics[scale=0.37]{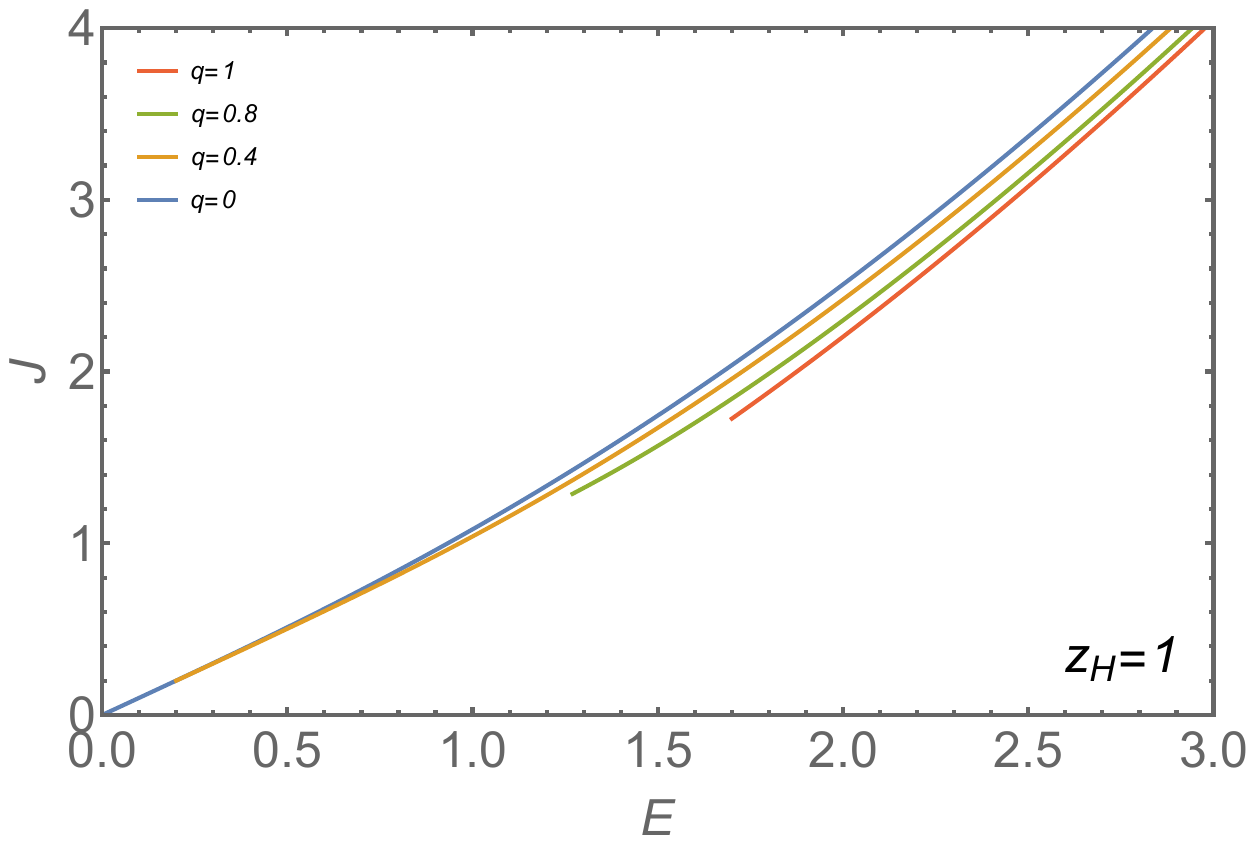}\includegraphics[scale=0.37]{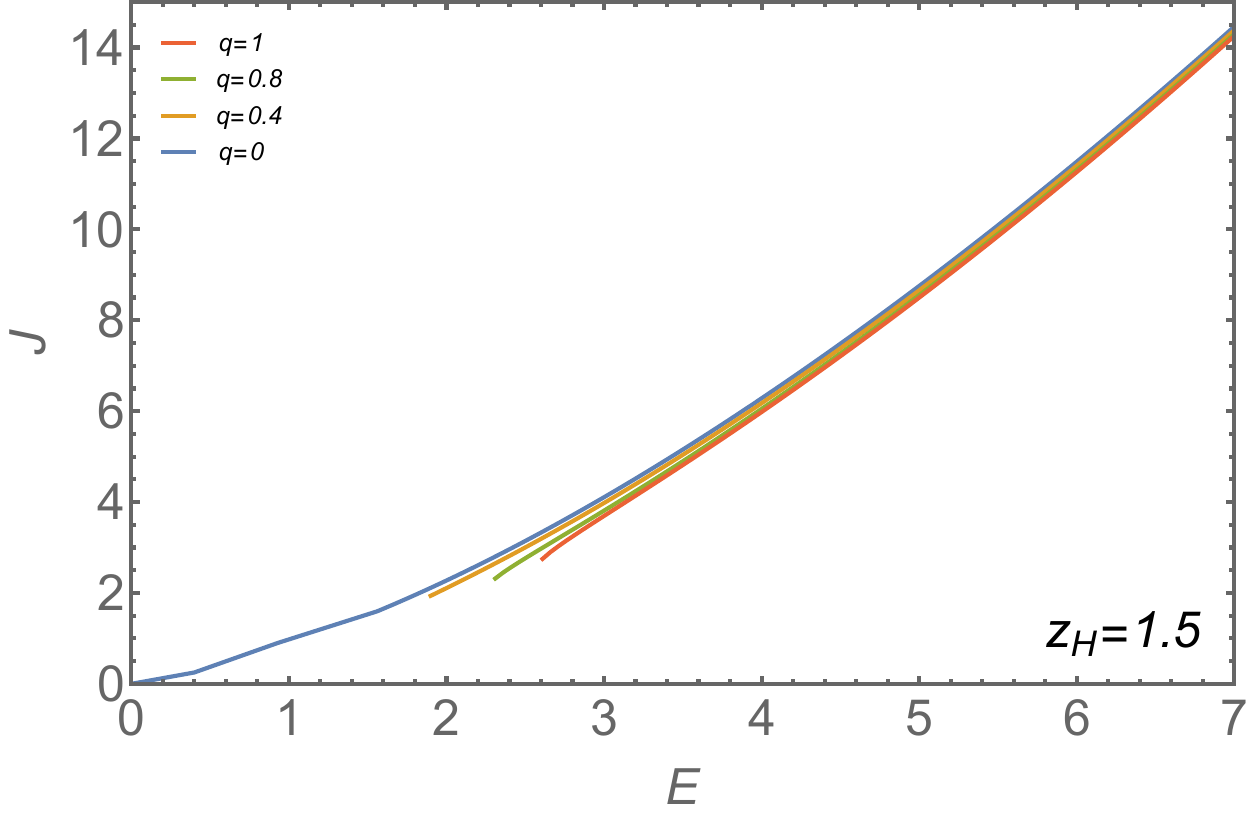}
\par\end{centering}
\caption{\label{fig:5}The vacuum V-A curve in holography with vanished magnetic
field. The stable current becomes non-zero at $E>E_{c}$. }

\end{figure}

In this section, we study the V-A curve of the instantonic vacuum
by using the effective Lagrangian (\ref{eq:2.10}), thus the electric
charge and magnetic field can be turned off as $d=0,B_{i}=0$ for
simplicity. The V-A curve comes from the relation of electric field
$E>E_{c}$ and stable current $J$ which corresponds to the reality
condition of the D-brane action \cite{Karch:2007pd,Karch:2007br,Erdmenger:2007bn}.
That means the D-brane configuration must be stable which does not
admit an imaginary part of the action. To this goal, we follow the
discussion in \cite{Hashimoto:2013mua,Hashimoto:2014dza}, that means
there is a certain position $z=z_{p}$ where the denominator of $\xi$
changes its sign, and at the same position $z=z_{p}$ the numerator
of $\xi$ changes its sign as well. Hence $z_{p}$ must be determined
by the equation as,

\begin{equation}
\left[1-\frac{\left(2\pi\alpha^{\prime}\right)^{2}z^{4}}{R^{4}}e^{-\phi}\left(E_{1}f^{-1}-B^{2}\right)-\frac{\left(2\pi\alpha^{\prime}\right)^{4}z^{8}}{R^{8}}e^{-2\phi}E_{1}^{2}B_{1}^{2}f^{-1}\right]\big|_{z=z_{p}}=0,\label{eq:3.4}
\end{equation}
for any given $E_{i},B_{i}$. And the corresponding stable current
$J$ can be determined by solving its denominator given by

\begin{equation}
\left[1-\frac{z^{6}j^{2}f^{-1}}{e^{\phi}R^{4}+\left(2\pi\alpha^{\prime}\right)^{2}B_{1}^{2}z^{4}}\right]\big|_{z=z_{p}}=0,\label{eq:3.5}
\end{equation}
with $d=0$. The numerical solution to (\ref{eq:3.4}) and (\ref{eq:3.5})
as the relation between $E$ and $J$ is given in Figure \ref{fig:5}.
 We can see the stable current becomes non-zero when the electric
field as an external field is larger than the critical value. Thus
the associated conductivity illustrates the vacuum is quite insulated
in the presence of instanton if the external electric field is not
sufficiently strong. The reason is that the instanton in our D(-1)-D3
system leads to a self-dual gauge field background \cite{Liu:1999fc}
as the vacuum in which the instanton trends to be neutral. And this
is also consistent with the behavior of V-A curve at large $E$ since
the conductivity is fairly insensitive with the instanton charge as
it is illustrated in Figure \ref{fig:5}.

\section{Summary}

In this work, we investigate the electromagnetic instability of the
instantonic vacuum by using the D(-1)-D3-brane system through gauge-gravity
duality. Since the D(-1)-D3-brane system describes the instantonic
plasma in holography, we consider the action for the D7-brane, as
the effective flavored action, in order to evaluate the electromagnetic
instability with instantons in the plasma. Our numerical calculation
illustrates the critical electric field increases rapidly as the instanton
density grows, and it leads to a maximum value of the vacuum decay
rate for the given external fields. To confirm this result, we further
derive the formulas of the critical electric field and find these
features correspond to the topological property of the instanton.
Since the instanton density increases the chiral condensate, particles,
as quarks or mesons, acquire the effective mass from the chiral condensate
as it is discussed in the QFT \cite{Gross:1980br,Schafer:1996wv,Vicari:2008jw}.
So the critical electric field must match to the mass of the particles
which is affected by the density of the instanton, otherwise the vacuum
decay does not occur. In addition, the V-A curve is also investigated
with instantons in this work which also reveals the electric current
does exist if the electric field is larger than a critical value.
In this sense, it implies the vacuum trends to be insulating with
instantons and the vacuum decay can occur only through a tunneling
process when the electromagnetic field strength is large. Overall,
this work illustrates insulating/conductive phase transition of the
instantonic vacuum with respect to the instanton density, and the
relation between electromagnetic and instantonic properties of the
vacuum in the plasma, as an extension of the existing works in the
D(-1)-D3 system \cite{Li:2020azb,Shahkarami:2015qff}\footnote{This work is also an extension of the D0-D4 approach with respect
to the analysis of the vacuum electromagnetic instability\cite{Cai:2016jgr}. }.

\section*{Acknowledgements}

The authors are supported by the National Natural Science Foundation
of China (NSFC) under Grant No. 12005033 and 11705026, the Fundamental
Research Funds for the Central Universities under Grant No. 3132025200
and 3132020178.

\bibliographystyle{utphys}
\bibliography{Electromagnetic_instability_of_vacuum_in_holographic_plasma_with_instanton}

\providecommand{\href}[2]{#2}\begingroup\raggedright\begin{thebibliography}{10}

\bibitem{Gross:1980br}
D.~J. Gross, R.~D. Pisarski, and L.~G. Yaffe, ``{QCD and Instantons at Finite
  Temperature},'' \href{http://dx.doi.org/10.1103/RevModPhys.53.43}{{\em Rev.
  Mod. Phys.} {\bfseries 53} (1981) 43}.

\bibitem{Schafer:1996wv}
T.~Sch{\"a}fer and E.~V. Shuryak, ``{Instantons in QCD},''
  \href{http://dx.doi.org/10.1103/RevModPhys.70.323}{{\em Rev. Mod. Phys.}
  {\bfseries 70} (1998) 323--426},
  \href{http://arxiv.org/abs/hep-ph/9610451}{{\ttfamily arXiv:hep-ph/9610451}}.

\bibitem{Vicari:2008jw}
E.~Vicari and H.~Panagopoulos, ``{Theta dependence of SU(N) gauge theories in
  the presence of a topological term},''
  \href{http://dx.doi.org/10.1016/j.physrep.2008.10.001}{{\em Phys. Rept.}
  {\bfseries 470} (2009) 93--150},
  \href{http://arxiv.org/abs/0803.1593}{{\ttfamily arXiv:0803.1593 [hep-th]}}.

\bibitem{Diakonov:2004jn}
D.~Diakonov, N.~Gromov, V.~Petrov, and S.~Slizovskiy, ``{Quantum weights of
  dyons and of instantons with nontrivial holonomy},''
  \href{http://dx.doi.org/10.1103/PhysRevD.70.036003}{{\em Phys. Rev. D}
  {\bfseries 70} (2004) 036003},
  \href{http://arxiv.org/abs/hep-th/0404042}{{\ttfamily arXiv:hep-th/0404042}}.

\bibitem{Diakonov:2007nv}
D.~Diakonov and V.~Petrov, ``{Confining ensemble of dyons},''
  \href{http://dx.doi.org/10.1103/PhysRevD.76.056001}{{\em Phys. Rev. D}
  {\bfseries 76} (2007) 056001},
  \href{http://arxiv.org/abs/0704.3181}{{\ttfamily arXiv:0704.3181 [hep-th]}}.

\bibitem{Liu:2015ufa}
Y.~Liu, E.~Shuryak, and I.~Zahed, ``{Confining dyon-antidyon Coulomb liquid
  model. I.},'' \href{http://dx.doi.org/10.1103/PhysRevD.92.085006}{{\em Phys.
  Rev. D} {\bfseries 92} no.~8, (2015) 085006},
  \href{http://arxiv.org/abs/1503.03058}{{\ttfamily arXiv:1503.03058
  [hep-ph]}}.

\bibitem{Liu:2015jsa}
Y.~Liu, E.~Shuryak, and I.~Zahed, ``{Light quarks in the screened dyon-antidyon
  Coulomb liquid model. II.},''
  \href{http://dx.doi.org/10.1103/PhysRevD.92.085007}{{\em Phys. Rev. D}
  {\bfseries 92} no.~8, (2015) 085007},
  \href{http://arxiv.org/abs/1503.09148}{{\ttfamily arXiv:1503.09148
  [hep-ph]}}.

\bibitem{Liu:2016thw}
Y.~Liu, E.~Shuryak, and I.~Zahed, ``{The Instanton-Dyon Liquid Model III:
  Finite Chemical Potential},''
  \href{http://dx.doi.org/10.1103/PhysRevD.94.105011}{{\em Phys. Rev. D}
  {\bfseries 94} no.~10, (2016) 105011},
  \href{http://arxiv.org/abs/1606.07009}{{\ttfamily arXiv:1606.07009
  [hep-ph]}}.

\bibitem{Liu:2016mrk}
Y.~Liu, E.~Shuryak, and I.~Zahed, ``{Light Adjoint Quarks in the Instanton-Dyon
  Liquid Model IV},'' \href{http://dx.doi.org/10.1103/PhysRevD.94.105012}{{\em
  Phys. Rev. D} {\bfseries 94} no.~10, (2016) 105012},
  \href{http://arxiv.org/abs/1605.07584}{{\ttfamily arXiv:1605.07584
  [hep-ph]}}.

\bibitem{Liu:2016yij}
Y.~Liu, E.~Shuryak, and I.~Zahed, ``{The Instanton-Dyon Liquid Model V: Twisted
  Light Quarks},'' \href{http://dx.doi.org/10.1103/PhysRevD.94.105013}{{\em
  Phys. Rev. D} {\bfseries 94} no.~10, (2016) 105013},
  \href{http://arxiv.org/abs/1606.02996}{{\ttfamily arXiv:1606.02996
  [hep-ph]}}.

\bibitem{Poppitz:2012nz}
E.~Poppitz, T.~Sch{\"a}fer, and M.~{\"U}nsal, ``{Universal mechanism of
  (semi-classical) deconfinement and theta-dependence for all simple groups},''
  \href{http://dx.doi.org/10.1007/JHEP03(2013)087}{{\em JHEP} {\bfseries 03}
  (2013) 087}, \href{http://arxiv.org/abs/1212.1238}{{\ttfamily arXiv:1212.1238
  [hep-th]}}.

\bibitem{Poppitz:2012sw}
E.~Poppitz, T.~Sch{\"a}fer, and M.~Unsal, ``{Continuity, Deconfinement, and
  (Super) Yang-Mills Theory},''
  \href{http://dx.doi.org/10.1007/JHEP10(2012)115}{{\em JHEP} {\bfseries 10}
  (2012) 115}, \href{http://arxiv.org/abs/1205.0290}{{\ttfamily arXiv:1205.0290
  [hep-th]}}.

\bibitem{Tong:2005un}
D.~Tong, ``{TASI lectures on solitons: Instantons, monopoles, vortices and
  kinks},'' in {\em {Theoretical Advanced Study Institute in Elementary
  Particle Physics}: {Many Dimensions of String Theory}}.
\newblock 6, 2005.
\newblock \href{http://arxiv.org/abs/hep-th/0509216}{{\ttfamily
  arXiv:hep-th/0509216}}.

\bibitem{HFLAV:2022esi}
{\bfseries HFLAV} Collaboration, Y.~S. Amhis {\em et~al.}, ``{Averages of
  b-hadron, c-hadron, and {\ensuremath{\tau}}-lepton properties as of 2021},''
  \href{http://dx.doi.org/10.1103/PhysRevD.107.052008}{{\em Phys. Rev. D}
  {\bfseries 107} no.~5, (2023) 052008},
  \href{http://arxiv.org/abs/2206.07501}{{\ttfamily arXiv:2206.07501
  [hep-ex]}}.

\bibitem{LHCb:2025ray}
{\bfseries LHCb} Collaboration, R.~Aaij {\em et~al.}, ``{Observation of
  charge{\textendash}parity symmetry breaking in baryon decays},''
  \href{http://dx.doi.org/10.1038/s41586-025-09119-3}{{\em Nature} {\bfseries
  643} no.~8074, (2025) 1223--1228},
  \href{http://arxiv.org/abs/2503.16954}{{\ttfamily arXiv:2503.16954
  [hep-ex]}}.

\bibitem{Maldacena:1997re}
J.~M. Maldacena, ``{The Large $N$ limit of superconformal field theories and
  supergravity},'' \href{http://dx.doi.org/10.4310/ATMP.1998.v2.n2.a1}{{\em
  Adv. Theor. Math. Phys.} {\bfseries 2} (1998) 231--252},
  \href{http://arxiv.org/abs/hep-th/9711200}{{\ttfamily arXiv:hep-th/9711200}}.

\bibitem{Aharony:1999ti}
O.~Aharony, S.~S. Gubser, J.~M. Maldacena, H.~Ooguri, and Y.~Oz, ``{Large N
  field theories, string theory and gravity},''
  \href{http://dx.doi.org/10.1016/S0370-1573(99)00083-6}{{\em Phys. Rept.}
  {\bfseries 323} (2000) 183--386},
  \href{http://arxiv.org/abs/hep-th/9905111}{{\ttfamily arXiv:hep-th/9905111}}.

\bibitem{Witten:1998qj}
E.~Witten, ``{Anti de Sitter space and holography},''
  \href{http://dx.doi.org/10.4310/ATMP.1998.v2.n2.a2}{{\em Adv. Theor. Math.
  Phys.} {\bfseries 2} (1998) 253--291},
  \href{http://arxiv.org/abs/hep-th/9802150}{{\ttfamily arXiv:hep-th/9802150}}.

\bibitem{Liu:1999fc}
H.~Liu and A.~A. Tseytlin, ``{D3-brane D instanton configuration and N=4
  superYM theory in constant selfdual background},''
  \href{http://dx.doi.org/10.1016/S0550-3213(99)00259-X}{{\em Nucl. Phys. B}
  {\bfseries 553} (1999) 231--249},
  \href{http://arxiv.org/abs/hep-th/9903091}{{\ttfamily arXiv:hep-th/9903091}}.

\bibitem{Casalderrey-Solana:2011dxg}
J.~Casalderrey-Solana, H.~Liu, D.~Mateos, K.~Rajagopal, and U.~Achim~Wiedemann,
  \href{http://dx.doi.org/10.1017/9781009403504}{{\em {Gauge/String Duality,
  Hot QCD and Heavy Ion Collisions}}}.
\newblock Cambridge University Press, 2014.
\newblock \href{http://arxiv.org/abs/1101.0618}{{\ttfamily arXiv:1101.0618
  [hep-th]}}.

\bibitem{Gwak:2012ht}
B.~Gwak, M.~Kim, B.-H. Lee, Y.~Seo, and S.-J. Sin, ``{Holographic D Instanton
  Liquid and chiral transition},''
  \href{http://dx.doi.org/10.1103/PhysRevD.86.026010}{{\em Phys. Rev. D}
  {\bfseries 86} (2012) 026010},
  \href{http://arxiv.org/abs/1203.4883}{{\ttfamily arXiv:1203.4883 [hep-th]}}.

\bibitem{Chen:2022obe}
J.-X. Chen and D.-F. Hou, ``{Heavy quark potential and jet quenching parameter
  in a rotating D-instanton background},''
  \href{http://dx.doi.org/10.1140/epjc/s10052-024-12708-7}{{\em Eur. Phys. J.
  C} {\bfseries 84} no.~4, (2024) 447},
  \href{http://arxiv.org/abs/2202.00888}{{\ttfamily arXiv:2202.00888
  [hep-ph]}}.

\bibitem{Li:2017ywp}
S.-w. Li and S.~Lin, ``{D-instantons in Real Time Dynamics},''
  \href{http://dx.doi.org/10.1103/PhysRevD.98.066002}{{\em Phys. Rev. D}
  {\bfseries 98} no.~6, (2018) 066002},
  \href{http://arxiv.org/abs/1711.06365}{{\ttfamily arXiv:1711.06365
  [hep-th]}}.

\bibitem{Li:2025ahp}
S.-w. Li and X.-t. Zhang, ``{Worldvolume fermion as baryon with homogeneous
  instantons in holographic QCD3},''
  \href{http://dx.doi.org/10.1103/d1p6-4yl9}{{\em Phys. Rev. D} {\bfseries 112}
  no.~2, (2025) 026001}, \href{http://arxiv.org/abs/2503.24173}{{\ttfamily
  arXiv:2503.24173 [hep-th]}}.

\bibitem{Li:2015kma}
S.-w. Li and T.~Jia, ``{Matrix model and Holographic Baryons in the D0-D4
  background},'' \href{http://dx.doi.org/10.1103/PhysRevD.92.046007}{{\em Phys.
  Rev. D} {\bfseries 92} no.~4, (2015) 046007},
  \href{http://arxiv.org/abs/1506.00068}{{\ttfamily arXiv:1506.00068
  [hep-th]}}.

\bibitem{Li:2015oza}
S.-w. Li, ``{Glueball{\textendash}baryon interactions in holographic QCD},''
  \href{http://dx.doi.org/10.1016/j.physletb.2017.08.011}{{\em Phys. Lett. B}
  {\bfseries 773} (2017) 142--149},
  \href{http://arxiv.org/abs/1509.06914}{{\ttfamily arXiv:1509.06914
  [hep-th]}}.

\bibitem{Li:2016kfo}
S.-w. Li and T.~Jia, ``{Three-body force for baryons from the D0-D4/D8 brane
  matrix model},'' \href{http://dx.doi.org/10.1103/PhysRevD.93.065051}{{\em
  Phys. Rev. D} {\bfseries 93} no.~6, (2016) 065051},
  \href{http://arxiv.org/abs/1602.02259}{{\ttfamily arXiv:1602.02259
  [hep-th]}}.

\bibitem{Li:2024apc}
S.-w. Li, Y.-p. Zhang, and H.-q. Li, ``{Holographic spectroscopy of fermion
  with instantons},''
  \href{http://dx.doi.org/10.1140/epjc/s10052-025-14556-5}{{\em Eur. Phys. J.
  C} {\bfseries 85} no.~8, (2025) 830},
  \href{http://arxiv.org/abs/2406.11557}{{\ttfamily arXiv:2406.11557
  [hep-th]}}.

\bibitem{Li:2021vve}
S.-w. Li, S.-k. Luo, and M.-z. Tan, ``{Three-dimensional
  Yang-Mills-Chern-Simons theory from a D3-brane background with
  D-instantons},'' \href{http://dx.doi.org/10.1103/PhysRevD.104.066008}{{\em
  Phys. Rev. D} {\bfseries 104} no.~6, (2021) 066008},
  \href{http://arxiv.org/abs/2106.04038}{{\ttfamily arXiv:2106.04038
  [hep-th]}}.

\bibitem{Li:2022wwv}
S.-w. Li, S.-k. Luo, and Y.-q. Hu, ``{Holographic QCD$_{3}$ and Chern-Simons
  theory from anisotropic supergravity},''
  \href{http://dx.doi.org/10.1007/JHEP06(2022)040}{{\em JHEP} {\bfseries 06}
  (2022) 040}, \href{http://arxiv.org/abs/2203.14489}{{\ttfamily
  arXiv:2203.14489 [hep-th]}}.

\bibitem{Schwinger:1951nm}
J.~S. Schwinger, ``{On gauge invariance and vacuum polarization},''
  \href{http://dx.doi.org/10.1103/PhysRev.82.664}{{\em Phys. Rev.} {\bfseries
  82} (1951) 664--679}.

\bibitem{Affleck:1981bma}
I.~K. Affleck, O.~Alvarez, and N.~S. Manton, ``{Pair Production at Strong
  Coupling in Weak External Fields},''
  \href{http://dx.doi.org/10.1016/0550-3213(82)90455-2}{{\em Nucl. Phys. B}
  {\bfseries 197} (1982) 509--519}.

\bibitem{Buckley:1999mv}
K.~Buckley, T.~Fugleberg, and A.~Zhitnitsky, ``{Can theta vacua be created in
  heavy ion collisions?},''
  \href{http://dx.doi.org/10.1103/PhysRevLett.84.4814}{{\em Phys. Rev. Lett.}
  {\bfseries 84} (2000) 4814--4817},
  \href{http://arxiv.org/abs/hep-ph/9910229}{{\ttfamily arXiv:hep-ph/9910229}}.

\bibitem{Rotter:2009zhs}
I.~Rotter, ``{A non-Hermitian Hamilton operator and the physics of open quantum
  systems},'' \href{http://dx.doi.org/10.1088/1751-8113/42/15/153001}{{\em J.
  Phys. A} {\bfseries 42} no.~15, (2009) 153001}.

\bibitem{Ge:2019crj}
Z.-Y. Ge, Y.-R. Zhang, T.~Liu, S.-W. Li, H.~Fan, and F.~Nori, ``{Topological
  band theory for non-Hermitian systems from the Dirac equation},''
  \href{http://dx.doi.org/10.1103/PhysRevB.100.054105}{{\em Phys. Rev. B}
  {\bfseries 100} no.~5, (2019) 054105},
  \href{http://arxiv.org/abs/1903.09985}{{\ttfamily arXiv:1903.09985
  [cond-mat.mes-hall]}}.

\bibitem{Shen:2018cjc}
H.~Shen, B.~Zhen, and L.~Fu, ``{Topological Band Theory for Non-Hermitian
  Hamiltonians},'' \href{http://dx.doi.org/10.1103/PhysRevLett.120.146402}{{\em
  Phys. Rev. Lett.} {\bfseries 120} no.~14, (2018) 146402}.

\bibitem{Hashimoto:2013mua}
K.~Hashimoto and T.~Oka, ``{Vacuum Instability in Electric Fields via AdS/CFT:
  Euler-Heisenberg Lagrangian and Planckian Thermalization},''
  \href{http://dx.doi.org/10.1007/JHEP10(2013)116}{{\em JHEP} {\bfseries 10}
  (2013) 116}, \href{http://arxiv.org/abs/1307.7423}{{\ttfamily arXiv:1307.7423
  [hep-th]}}.

\bibitem{Hashimoto:2014dza}
K.~Hashimoto, T.~Oka, and A.~Sonoda, ``{Magnetic instability in AdS/CFT:
  Schwinger effect and Euler-Heisenberg Lagrangian of supersymmetric QCD},''
  \href{http://dx.doi.org/10.1007/JHEP06(2014)085}{{\em JHEP} {\bfseries 06}
  (2014) 085}, \href{http://arxiv.org/abs/1403.6336}{{\ttfamily arXiv:1403.6336
  [hep-th]}}.

\bibitem{Li:2020azb}
S.-w. Li, ``{Holographic Schwinger effect in the confining background with
  D-instanton},'' \href{http://dx.doi.org/10.1140/epjc/s10052-021-09607-6}{{\em
  Eur. Phys. J. C} {\bfseries 81} no.~9, (2021) 797},
  \href{http://arxiv.org/abs/2005.11955}{{\ttfamily arXiv:2005.11955
  [hep-th]}}.

\bibitem{Shahkarami:2015qff}
L.~Shahkarami, M.~Dehghani, and P.~Dehghani, ``{Holographic Schwinger Effect in
  a D-Instanton Background},''
  \href{http://dx.doi.org/10.1103/PhysRevD.97.046013}{{\em Phys. Rev. D}
  {\bfseries 97} no.~4, (2018) 046013},
  \href{http://arxiv.org/abs/1511.07986}{{\ttfamily arXiv:1511.07986
  [hep-th]}}.

\bibitem{Karch:2002sh}
A.~Karch and E.~Katz, ``{Adding flavor to AdS / CFT},''
  \href{http://dx.doi.org/10.1088/1126-6708/2002/06/043}{{\em JHEP} {\bfseries
  06} (2002) 043}, \href{http://arxiv.org/abs/hep-th/0205236}{{\ttfamily
  arXiv:hep-th/0205236}}.

\bibitem{Semenoff:2011ng}
G.~W. Semenoff and K.~Zarembo, ``{Holographic Schwinger Effect},''
  \href{http://dx.doi.org/10.1103/PhysRevLett.107.171601}{{\em Phys. Rev.
  Lett.} {\bfseries 107} (2011) 171601},
  \href{http://arxiv.org/abs/1109.2920}{{\ttfamily arXiv:1109.2920 [hep-th]}}.

\bibitem{Karch:2007pd}
A.~Karch and A.~O'Bannon, ``{Metallic AdS/CFT},''
  \href{http://dx.doi.org/10.1088/1126-6708/2007/09/024}{{\em JHEP} {\bfseries
  09} (2007) 024}, \href{http://arxiv.org/abs/0705.3870}{{\ttfamily
  arXiv:0705.3870 [hep-th]}}.

\bibitem{Karch:2007br}
A.~Karch and A.~O'Bannon, ``{Holographic thermodynamics at finite baryon
  density: Some exact results},''
  \href{http://dx.doi.org/10.1088/1126-6708/2007/11/074}{{\em JHEP} {\bfseries
  11} (2007) 074}, \href{http://arxiv.org/abs/0709.0570}{{\ttfamily
  arXiv:0709.0570 [hep-th]}}.

\bibitem{Erdmenger:2007bn}
J.~Erdmenger, R.~Meyer, and J.~P. Shock, ``{AdS/CFT with flavour in electric
  and magnetic Kalb-Ramond fields},''
  \href{http://dx.doi.org/10.1088/1126-6708/2007/12/091}{{\em JHEP} {\bfseries
  12} (2007) 091}, \href{http://arxiv.org/abs/0709.1551}{{\ttfamily
  arXiv:0709.1551 [hep-th]}}.

\bibitem{Cai:2016jgr}
W.~Cai, K.-l. Li, and S.-w. Li, ``{Electromagnetic instability and Schwinger
  effect in the Witten{\textendash}Sakai{\textendash}Sugimoto model with
  D0{\textendash}D4 background},''
  \href{http://dx.doi.org/10.1140/epjc/s10052-019-7404-1}{{\em Eur. Phys. J. C}
  {\bfseries 79} no.~11, (2019) 904},
  \href{http://arxiv.org/abs/1612.07087}{{\ttfamily arXiv:1612.07087
  [hep-th]}}.

\end{thebibliography}\endgroup

\end{document}